\documentclass[sn-standardnature]{sn-jnl}% Style for submissions to Nature Portfolio journals
\usepackage{graphicx}%
\usepackage{multirow}%
\usepackage{amsmath,amssymb,amsfonts}%
\usepackage{amsthm}%
\usepackage{mathrsfs}%
\usepackage[title]{appendix}%
\usepackage{xcolor}%
\usepackage{textcomp}%
\usepackage{manyfoot}%
\usepackage{booktabs}%
\usepackage{algorithm}%
\usepackage{algorithmicx}%
\usepackage{algpseudocode}%
\usepackage{listings}%
\usepackage[utf8]{inputenc}
\theoremstyle{thmstyleone}%
\theoremstyle{thmstyletwo}%

\theoremstyle{thmstylethree}%

\begin{document}

\title[Increased H escape in recent Mars]{Increased Hydrogen escape from Mars atmosphere  during periods of high obliquity}

%%=============================================================%%
%% GivenName	-> \fnm{Joergen W.}
%% Particle	-> \spfx{van der} -> surname prefix
%% FamilyName	-> \sur{Ploeg}
%% Suffix	-> \sfx{IV}
%% \author*[1,2]{\fnm{Joergen W.} \spfx{van der} \sur{Ploeg} 
%%  \sfx{IV}}\email{iauthor@gmail.com}
%%=============================================================%%

\author[1]{\fnm{Gabriella} \sur{Gilli}}

\author*[1]{\fnm{Francisco} \sur{Gonz\'alez-Galindo}}\email{ggalindo@iaa.es}

\author[2]{\fnm{Jean-Yves} \sur{Chaufray}}

\author[3]{\fnm{Ehouarn} \sur {Millour}}

\author[3]{\fnm{Fran\c cois} \sur {Forget}}

\author[2]{\fnm{Franck} \sur{Montmessin}}

\author[2]{\fnm{Franck} \sur{Lef\`evre}}

\author[3]{\fnm{Joseph} \sur {Naar}}

\author[3]{\fnm{Yangcheng} \sur{Luo}}

\author[2]{\fnm{Margaux} \sur{Vals}}

\author[2]{\fnm{Lo\"{\i}c} \sur{Rossi}}

\author[1]{\fnm{Miguel \'Angel} \sur{L\'opez-Valverde}}

\author[1]{\fnm{Adri\'an} \sur{Brines}}

\affil*[1]{\orgdiv{Instituto de Astrof\'{\i}sica de Andaluc\'{\i}a}, \orgname{CSIC}, \orgaddress{\street{Glorieta de la Astronom\'{\i}a}, \city{Granada}, \postcode{18008}, \country{Spain}}}

\affil[2]{\orgdiv{LATMOS/IPSL, UVSQ Université Paris-Saclay, Sorbonne Université}, \orgname{CNRS}, \orgaddress{ \city{Guyancourt}, \country{France}}}

\affil[3]{\orgdiv{Laboratoire de M\'et\'eorologie Dynamique}, \orgname{IPSL}, \orgname{Sorbonne Université, ENS, Université PSL, École Polytechnique, Institut Polytechnique, Institut Polytechnique de Paris}\orgaddress{\city{Paris},  \country{France}}}

%%==================================%%
%% sample for unstructured abstract %%
%%==================================%%

\abstract{\bf{It is still unknown how much water has escaped from Mars during its history. Hydrogen escape from Mars' atmosphere probably played a major role in drying the planet, but present-day H-loss rates ($\sim$3x10$^{26}$ s$^{-1}$ on average) cannot explain the geological evidence of the presence of large volumes of liquid water on ancient Mars. We use the 3D Mars-Planetary Climate Model to show that H-loss rates could have increased up to more than one order of magnitude (6x10$^{27}$ s$^{-1}$) during higher spin axis obliquity periods, notably in the last millions of years when Mars’s obliquity was about 35° on average. The resulting accumulated H escape over Mars history translates into  $\sim$ 80 m Global Equivalent Layer, which is close to the lower limit of geological estimates.}}


%%================================%%
%% Sample for structured abstract %%
%%================================%%

\maketitle
\section*{Main}\label{sec1}

Mars was not always as dry as it is today: several geologic and mineralogical observations indicate the evidence of past liquid water, with valley networks and outflow channels still visible on the surface \cite{Carr1987,Bibring2004}. Much of its surface water is believed to have been lost in early Mars history, but the water loss mechanisms and the amounts that escaped into space or remained frozen in the crust today are not well known. \iffalse \cite{Scheller2021} suggest that most of the water was lost between about 4.1 and 3.7 billion years, during the Noachian period, and one potential water loss mechanism is crustal hydration through irreversible chemical weathering, which could have sequestered between 30\%  and 90\%  of Martian water. The other proposed mechanism is water\fi Atmospheric thermal (Jeans) escape of atomic hydrogen (H) \cite{Anderson1971, McElroy1972, ParkinsonHunten1972} is considered as a dominant loss mechanism for water, at least in the last couple billion years \cite{Jakosky2018}. \iffalse The predicted dominant process for the H-escape is the thermal (Jeans) escape, derived from the photodissociation of water vapour followed by transport of H to the exobase ($\sim$ 200~km altitude), where at a temperature of $\sim$  200~K, a significant fraction of H atoms has enough speed to escape to space \cite{Clarke2018,Heavens2018}. \fi
Present-day loss rates of H by Jeans escape were determined by the MAVEN (Mars Atmosphere and Volatile EvolutioN) mission, with values varying with seasons from 1 to 11×10$^{26}$ atoms/s \cite{Jakosky2018, Halekas2017,Mayyasi2023}. Similar results were found by observations made from spacecraft like Mariner 6, 7, and 9 \cite{Anderson1974}, Mars Express (MEx) \cite{Chaffin2014}, Hubble Space Telescope (HST) \cite{Clarke2014} and Rosetta flyby of Mars \cite{Feldman2011}. %(see Fig. \ref{fig:H_obl_escape_compar}, bottom panel). 

Unexpected variations of the H corona were observed in the exosphere of Mars using the ultraviolet H Ly-alpha emission by MEx \cite{Chaffin2014} and HST \cite{Clarke2014}. A link between the observed variations and the enhanced propagation of water vapour at perihelion during a global dust storm was suggested. Subsequently, a 1D photochemical model \cite{Chaffin2017} described that fast transfer of H$_2$O to the upper atmosphere can produce fast enhancements of escape rates. This was later confirmed by observations from NGIMS (Neutral Gas and Ion Mass Spectrometer) on board MAVEN that showed how water is directly transported to the upper atmosphere of Mars during regional and global dust storms \cite{Stone2020}. The impact of regional dust storms on the middle atmosphere water content eventually resulting in enhanced H escape was also demonstrated by \cite{Heavens2018} and \cite{Chaffin2021} combining atmospheric retrievals from several spacecrafts: Mars Climate Sounder (MCS), Mars Reconnaissance Orbiter (MRO), MAVEN and Trace Gas Orbiter (TGO). All these results underlined the role of dust storms as an important factor allowing the penetration of water to the mesosphere and providing H-escape on Mars, although other studies showed  the regular seasonal variability in absence of discrete dust storms to be the dominant supplying process of escaping hydrogen \cite{Montmessin2022,kleinbohl2024}.

Altogether, a new paradigm has emerged in the last decade, supporting the hypothesis that direct intrusion of H$_2$O in the upper atmosphere is the dominant provider of H escape instead of the slow vertical transport of molecular hydrogen (H$_2$) \cite{Stone2020,Chaffin2021,Montmessin2022,kleinbohl2024}. \iffalse Therefore, the traditional picture of water escape was replaced by a new concept. In the classical scheme, water is chemically converted into molecular H (H$_2$) (with a long chemical lifetime) close to the surface, and slowly transported vertically to the thermosphere where it is converted into H atoms by ion-neutral reactions and leaves the atmosphere through H Jeans escape \cite{Stone2020}.\fi The new scenario takes into account the strong coupling with the processes and the state of the lower atmosphere, as mentioned above. 
Nevertheless, although the mechanism of water escape is now better understood, if we assume the estimated current H escape rates as representative of the whole Mars history, the accumulated water loss during the last 4 billion years is much lower (3-25 m, expressed in Global Equivalent Layer (GEL)) \cite{Jakosky2018,Chaufray2021,Chaffin2014} than the amount of water needed to form the flow channels observed on Mars ($\sim$ 100 to 1500 m GEL) \cite{Carr2003,Scheller2021}. 

\subsection*{Evidences of climate changes in recent Mars}
There are numerous evidences of climate changes in Mars in recent geological eras. Landforms resulting from the local accumulation of ice indicate that, even during the Amazonian period, large amounts of water (e.g. glaciers several hundred meters thick) formed and remained on the planet \cite{Forget2017}. These landforms reveal a different climate in which surface ice was mobilized and transferred across the planet during the last millions of years.

The climate of a planet depends on its orbital and rotation parameters, in particular on the obliquity (i.e., inclination of its spin axis relative to the perpendicular to its orbital plane). The primary cause of these climate changes could have been the variation of Mars' orbital and spin parameters \cite{Forget2017}. While for the Earth the oscillations of those parameters are small (± 1.3$^{\circ}$ for obliquity), in the past 250 Myrs Mars' axis inclination covered a large range of variations, between 0 and 66$^{\circ}$, with a mean obliquity of about 35$^{\circ}$ \cite{Laskar2004}. Different orbital configurations induce significant modifications in fundamental aspects of the Martian climate,  such as the CO$_2$ cycle (and thus the surface pressure), the dust and water cycle, or the global circulation \cite{Forget2017}, mainly due to the differences in the distribution of the insolation. 

The north pole insolation is a key parameter controlling the stability of water ice of the northern polar cap and in the past 10 million years it could have varied from about 140 W m$^{-2}$ to 400 W m$^{-2}$ for an obliquity of 15$^{\circ}$ and 35$^{\circ}$, respectively, resulting in a significant increase of the atmospheric water column \cite{Madeleine2014}. If part of this water was transferred to the upper atmosphere, a larger escape rate would be expected \cite{Jakosky2021}. This is consistent with estimations of larger escape in the past from the analysis of the D/H (deuterium to hydrogen) atmospheric enrichment \cite{Alsaeed2019,Cangi2020,Chaufray2024}. 

Previous theoretical studies focused on the impact of the obliquity changes in the lower atmosphere \cite{Madeleine2014}. However, to our knowledge, the effects of changes in the obliquity in the escape rate have not been quantified so far.

In this study, we aim to provide a first order estimation of this variation in past epochs characterised by changes in the obliquity during the last 10-20 million years of Mars history. We focus only on the effect of changing obliquity on the water cycle and the H escape rate, neglecting all other atmospheric effects, i.e., we assume that the bulk atmosphere at different obliquities remained the same as of today. This is equivalent to studying how current Mars would react to sudden changes in the obliquity and must be understood as a first, rough approximation aimed at isolating the effect of the obliquity on the escape rate.

\subsection*{3D simulations of current H escape \iffalse with high obliquity  \fi}\label{sec2}
We use an improved version of the 3D model used in \cite{Chaufray2021}, now called Mars-Planetary Climate Model (Mars-PCM), a 3D general circulation model of the Martian atmosphere and ionosphere extending from the surface to the exobase \cite{Chaufray2021,Vals2022, Rossi2022}. H-loss simulations for current conditions with a previous version of the Mars-PCM were performed by \cite{Chaufray2021} including for the first time the atmosphere-exosphere coupling. They confirmed the influence of episodic dust events on water enhancements at high altitude near the perihelion. Nevertheless, the amplitude of the variability observed by SPICAM/Mars Express was not reproduced by the Mars-PCM. We use in this study an updated version that presents two main differences with the one used in \cite{Chaufray2021}: the inclusion of a sophisticated microphysical model for water ice cloud formation \cite{Navarro2014}, and a larger set of photochemical reactions, including those involving water-derived ions \cite{Krasnopolsky2019,Chaufray2024}. These improvements are described in detail in the Methods Section. Only the H thermal (Jeans) escape is considered in this study. The thermal escape rate is computed using the same methodology as \cite{Chaufray2015} (i.e. considering the local vertical velocity of H at the upper boundary is equal to the local Jeans effusion velocity, and then integrating the flux over the surface).

\begin{center}
  \begin{figure}[htbp]
    \includegraphics[width=0.9\textwidth]{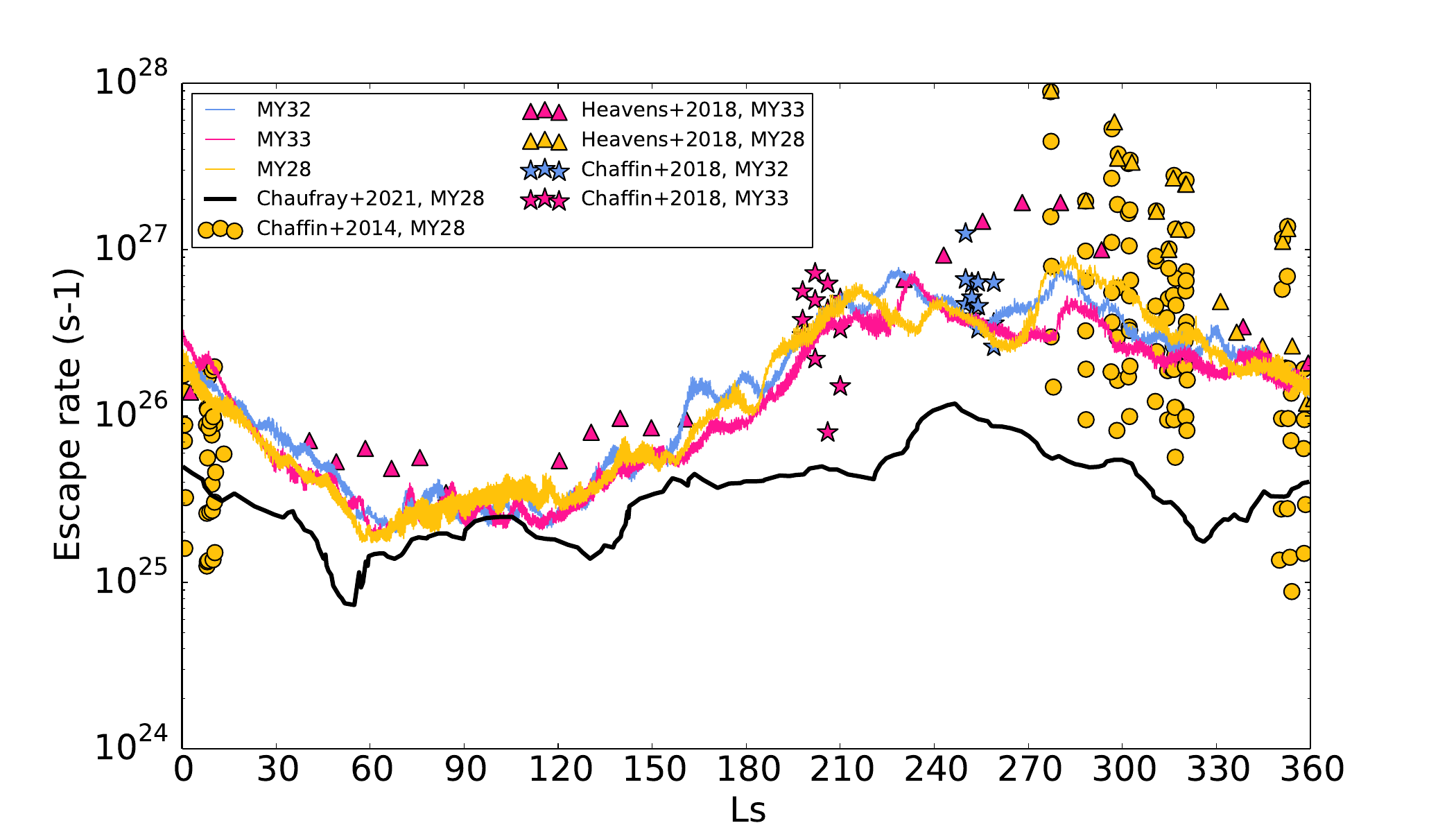}
    \caption{\textbf{Observed and modelled H escape rates (atoms/s) for current obliquity.} Globally integrated escape rate (atoms/s) simulated with the Mars-PCM. A simulation using a version similar to the one in \cite{Chaufray2021} is shown with a black line. Simulations with the improved version for different Mars Years are shown by colour lines as indicated in the legend. Different observational datasets are represented with symbols:  Mars Express observations during MY28 (yellow circles) \cite{Chaffin2014}, Hubble Space Telescope observations also taken during MY28 (yellow triangles) \cite{Heavens2018}, IUVS observations obtained during MY32 (blue stars) and MY33 (purple stars) \cite{Chaffin2018} and MAVEN SWIA observations during MY33 (purple triangles) \cite{Heavens2018}.}\label{fig:Hescape_compobs}
  \end{figure}
\end{center}

The impact of these improvements on our calculation of the the current H escape rate is shown in Fig. \ref{fig:Hescape_compobs}, comparing different observational datasets (see caption for more details) with the annual variation of the H escape rate simulated with a similar model version as in \cite{Chaufray2021} in black and the results of the improved model for three different Mars Years in color lines. While around the aphelion season the simulated H escape rate is similar to that in \cite{Chaufray2021}, the new model produces a significant increase in H escape rate from Ls~180° to Ls~60° (the year after), up to one order of magnitude larger than in the previous model version, particularly during the dusty season. This larger escape rate is in much better agreement with the observational datasets. Both the observed seasonal trend and the actual values are now well reproduced by the Mars-PCM. The average global H escape rate compares also very well with that obtained from the recent analysis of IUVS observations during MY32 to MY36 \cite{Mayyasi2023, Montabone2020}.
The various effects on the escape rate of the different improvements incorporated in the model are shown in Extended Data Fig. 1 and are discussed in depth in the Methods Section.

\section*{Results}
\subsection*{Higher loss rate in the “recent” Mars history}

\begin{center}
  \begin{figure}
  \centering
    \includegraphics[width=0.85\textwidth]{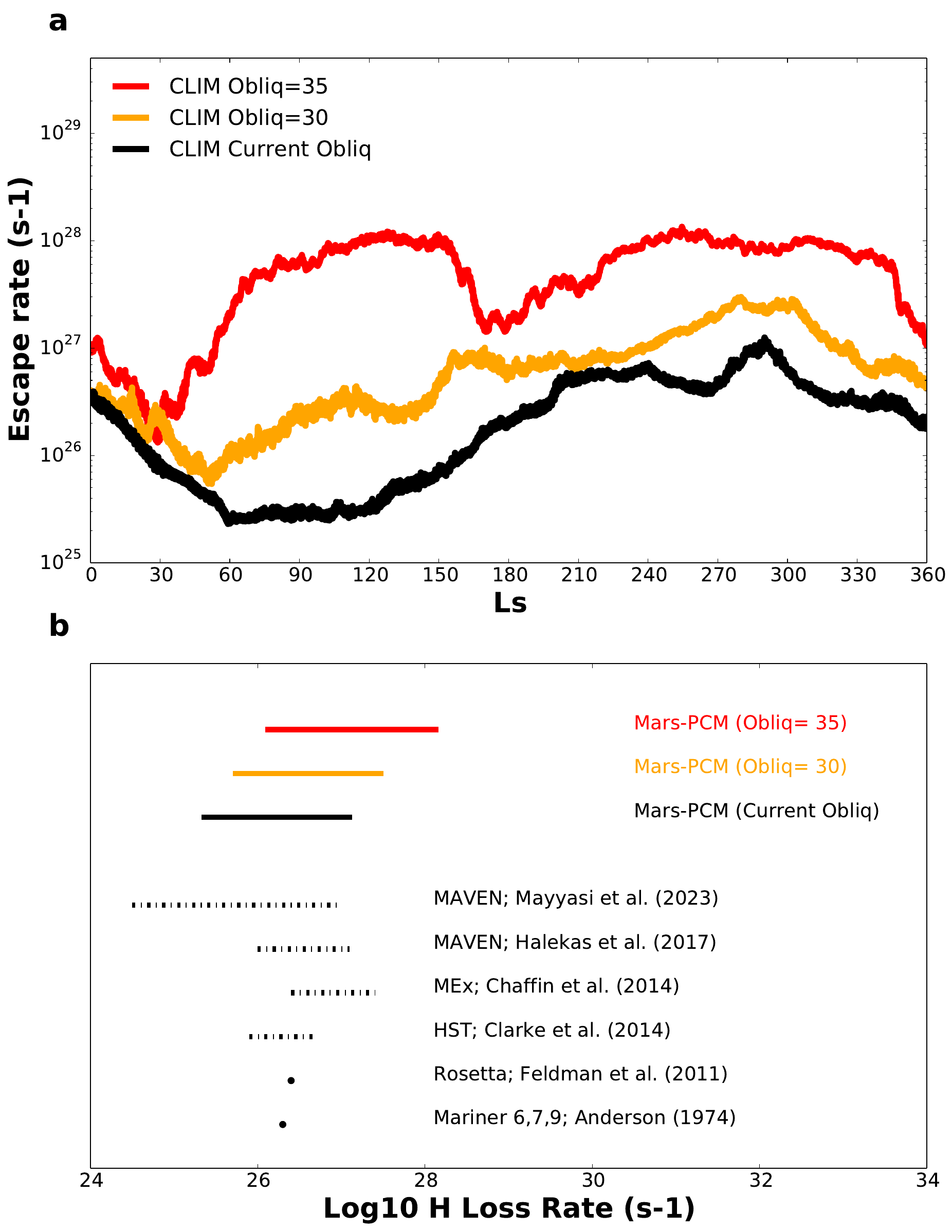}  
    \caption{\textbf{H escape rates simulated for different obliquity and from observations made for different spacecraft.} Panel a: Globally integrated escape rate (atoms/s) simulated with the Mars-PCM.  A climatological dust scenario called “CLIM” is used with current obliquity (black), obliquity of 30$^{\circ}$ (orange), obliquity of 35$^{\circ}$ (red). 
    Panel b: Comparison of Mars-PCM H escape rates with H-loss rates estimated for current obliquity from different spacecrafts \cite{Clarke2018,Heavens2018,Halekas2017,Mayyasi2023,Anderson1974, Chaffin2014, Clarke2014, Feldman2011}. }
  \label{fig:H_obl_escape_compar}
  \end{figure}
\end{center}

Fig. \ref{fig:H_obl_escape_compar} shows globally-integrated H escape rate,  for 3 different scenarios.  The reference climatology scenario with current obliquity (25.2$^{\circ}$) is in black, while the larger obliquities of 30$^{\circ}$ and 35$^{\circ}$ are orange and red, respectively. Obliquity 35$^{\circ}$ is common throughout Mars history, being a mean value within the past 20 My and quite close to the most probable value of 41.8$^{\circ}$ over 4 Gyrs \cite{Laskar2004}. 

We found that the H-loss is sensitive to the changes in the obliquity of Mars: the annually integrated H escape increases from 1 x 10$^{34}$ atoms in current conditions, to 1.9x10$^{35}$ for obliquity 35$^{\circ}$.
The maximum value of the global H-escape flux (i.e. the peak of the seasonal H-escape in a given year and for a given obliquity) varies from 8.92x10$^{26}$ atoms/s for current obliquity to about 1.17x10$^{28}$ atoms/s for high obliquity (35$^{\circ}$). Also the seasonal evolution of the H escape is different for the high obliquity simulation, with two periods of maximum escape around Northern summer and Northern winter, separated by a decrease in the escape rate. Our results indicate that a significant H loss could have taken place in the past, when the obliquity of Mars was higher than today. 

\subsection*{Processes leading to larger H-escape}

\begin{center}
  \begin{figure}[htbp]
    \centering
   \includegraphics[width=0.95\textwidth]{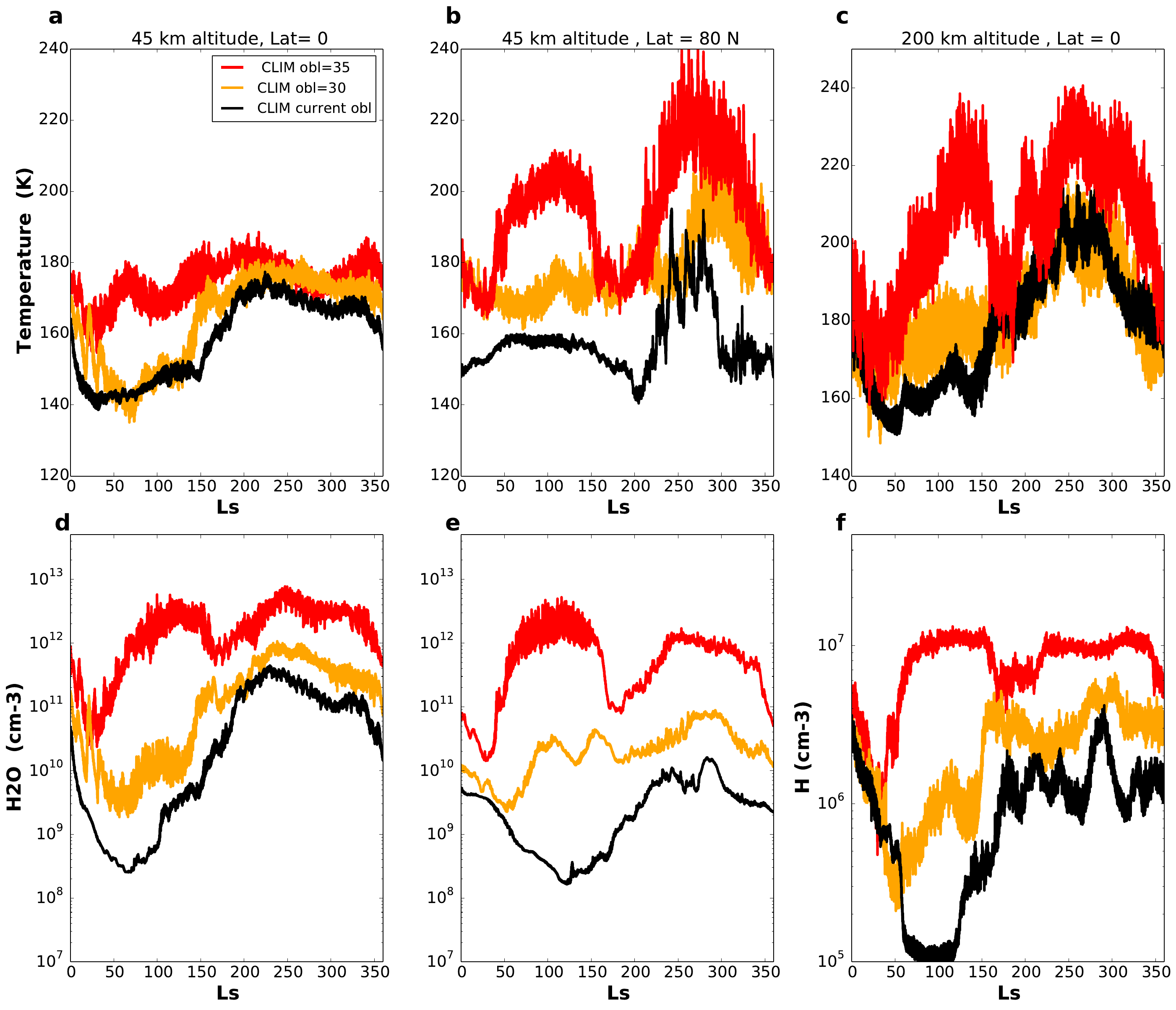}
\caption{\textbf{Seasonal variation of temperature, water vapour and atomic hydrogen number density for selected altitudes and latitudes during present-day and high obliquity cases.} Zonal averaged temperature and water vapour number density at 45 km, Latitude = 0 (panels a and d) and Latitude = 80$^{\circ}$ North (panels b and e). Temperature and H density at 200 km at Ls = 0 (panels c and f). The different obliquity scenarios are the same as in Fig.\ref{fig:H_obl_escape_compar}}
\label{fig:temp_water}
  \end{figure}
\end{center}

\begin{center}
  \begin{figure}[!htbp]
    \includegraphics[width=1\textwidth]{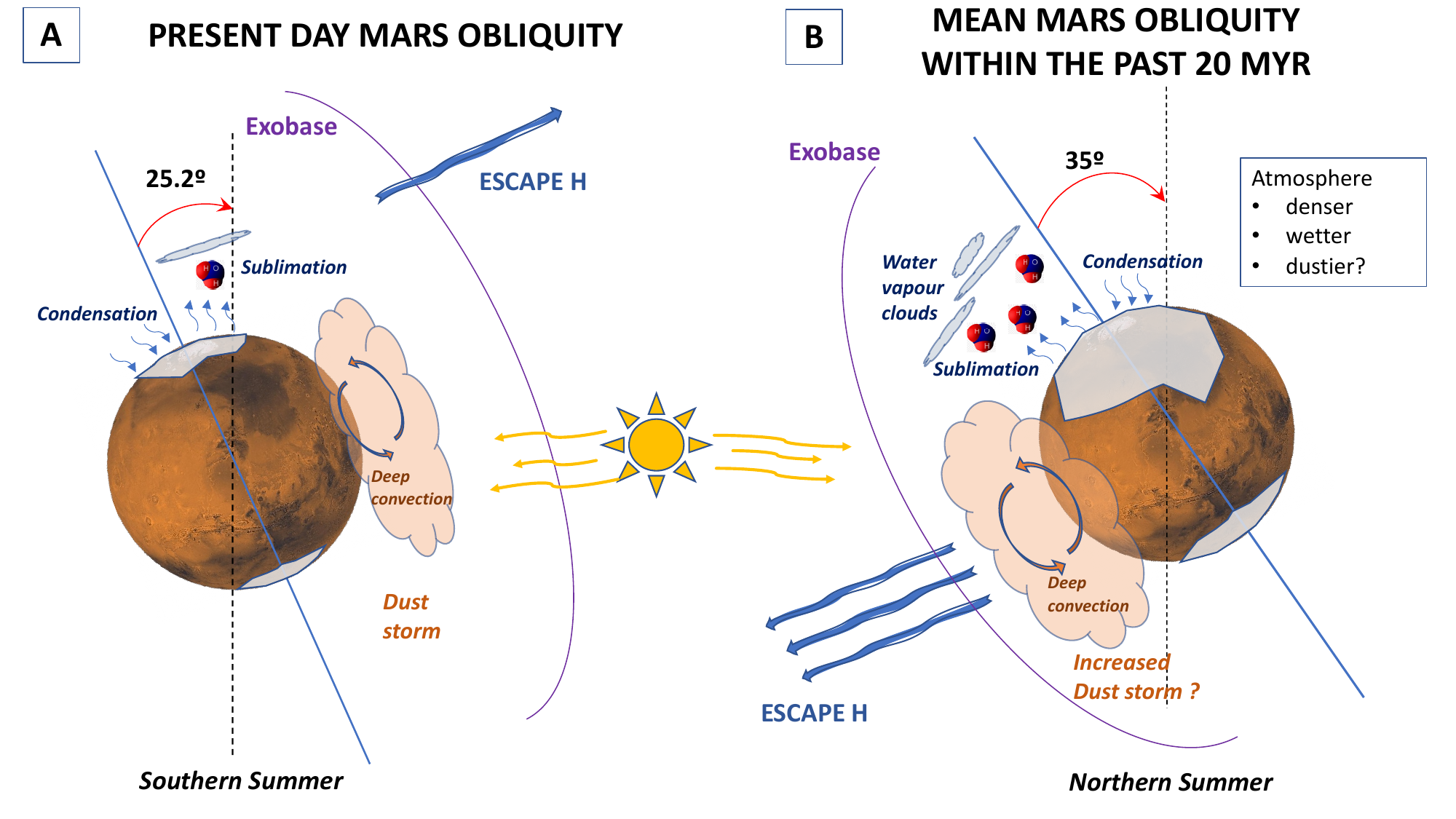}
  \caption{\textbf{Sketch showing processes leading to larger H-loss in the past 20 million years (Myr).} The loss rate is not expected to have been constant with time, and may vary significantly during Martian history. Panel A: In current obliquity conditions (25.2$^{\circ}$) the water ice in the polar caps sublimes in Summer, and then it is recycled back in Winter. Large dust load in the lower atmosphere facilitates the transport of water to the upper atmosphere, where it is chemically converted into atomic H that can easily escape to space. Panel B: In the last 20 million year, when the obliquity of Mars was higher than today, the climate of Mars was different. Larger north pole insolation induced a more intense water cycle: the amount of sublimated water vapour in the atmosphere of Mars was much larger than today, and localised surface water ice reservoirs were created after precipitation in tropics and mid-latitudes \cite{Madeleine2009}. In addition, the formation of thick clouds warmed the middle atmosphere (up to 50 K at 45 km) by absorbing both solar radiation and IR radiation emitted by the surface, inducing positive feedback. All this favoured water penetration into the mesosphere (e.g. with up to 5 order of magnitude increased water abundances at about 45 km, near the aphelion), resulting in larger H escape rate. Other processes not accounted for in our study could also contribute to further changes in the H escape rate. Buried deposits of CO$_2$ ice within the south polar layer could have been released in the atmosphere at the time of high obliquity, producing an atmosphere with double its current pressure \cite{Kurokawa2014}.
  With higher pressure and warmer temperature conditions, is uncertain if the seasonal dust activity was more or less intense than today, due to higher water content and changes in the circulation patterns. (Mars image credit: NASA/JPL/USGS).}
  %The model is unable to reliable simulate the effects of increased obliquity on the dust distribution in the atmosphere and can not be run enough to study the effect of increased pressure. Therefore, at this stage, we can only speculate about the additional effects of these processes on H escape, peaking in Southern Summer.
  \label{fig:sketch}
  \end{figure}
\end{center}

Changes in the obliquity of Mars in the last million years modified the mobilisation of water across the planet, notably affecting the water content of the atmosphere. In current conditions, the water ice sublimes in Summer, and it is recycled back to the polar caps in Winter. Variations in solar heating associated with orbital eccentricity can also produce large seasonal and hemispheric asymmetries in the martian climate system, therefore changing the deposition of water ice at the northern versus the southern pole. However, \cite{RichardsonWilson2002}  argued that the effect of $\sim45 \%$ increased insolation at periapsis is smaller than the North-to-South topographic dichotomy, which forces a dominant southern summer Hadley circulation and control the inter-hemispheric transport of water and dust. Previous studies with climate models at high obliquity \cite{Montmessin2006, Richardson2002, Haberle2011} showed a displacement of the water ice reservoirs outside the polar regions. More importantly, \cite{Richardson2002} showed that the increased polar summer insolation (e.g. for increased obliquities from 25$^{\circ}$ to 45$^{\circ}$) enhances the polar ice sublimation leading to a more intense water cycle (compared with today’s values) and higher column water abundances (up to 3,000 precipitable micrometres pr-$\mu$m vs. $\sim$70 pr-$\mu$m today) above the northern polar cap during summer solstice, and about 50 pr-$\mu$m in the summer tropics (vs. $\sim$10 pr-$\mu$m today). The effect is even more pronounced in simulations which take into account the radiative effect of clouds, as a more humid atmosphere results in the formation of thick water ice clouds that warm the atmosphere by absorbing both solar radiation and infrared radiation emitted by the surface \cite{Madeleine2014,Lasue2013}. Consequently, since the atmosphere can hold much more water vapour before saturation, thicker clouds could have formed, then inducing positive feedback. \cite{Madeleine2014} found an increase in the annual mean atmospheric temperature of $\sim$50 K in the tropics at an altitude of about 20 km when the obliquity is set to 35$^{\circ}$.
 %In our simulations with obliquity 35$^{\circ}$, higher water vapour column densities are found up to a factor 3 in polar summers (70º N/S) and up to a factor 2 at equatorial-mid latitudes (see Fig. \ref{fig:water_colum}).
Our simulations show that, in the middle atmosphere of Mars ($\sim$ 45 km), the warmer temperatures predicted for high obliquity (+40 K at low-mid latitudes and up to +110 K in the North polar region) contributed to the increase of the global water content at this altitude (up to 5 order of magnitude around the aphelion). As shown in Fig. \ref{fig:temp_water} the increase of temperature (and of water abundance) at those layers is particularly large during the first half of the year, due to the intense warming produced by the radiative effects of the thicker water ice clouds, as in \cite{Madeleine2014} (see Fig. \ref{fig:sketch}). This results in a change in the seasonal behavior for high obliquity, notably at mid-high latitudes, with two maxima of temperature and water abundance around both solstices, with minimum values around equinoxes. This behavior reflects clearly on both the H abundance at 200 km (see Fig. \ref{fig:temp_water} panel f) and on the H escape rate (Fig. \ref{fig:H_obl_escape_compar}).
 The thermospheric temperatures (at 200 km) are somehow less affected by the change in the obliquity, with little differences between the simulations for current obliquity and for obliquity 30$^{\circ}$, reflecting the dominance of the UV solar radiation input in setting the temperatures in the upper layers. However, the temperatures for the obliquity 35$^{\circ}$ case show also increases with respect to the current obliquity around both solstices, reflecting the much warmer atmosphere below, which would also contribute to  increasing the escape.
 This is not in contradiction with the topographically forced asymmetry in the martian circulation proposed in \cite{RichardsonWilson2002}. What our simulations show is that a 10$^\circ$ increase in obliquity is sufficient to release enough water in the Northern Summer and be advected to higher altitude (thanks also to the higher temperature) to compensate for the weaker Hadley circulation of the Northern summer/spring.

A simulation for obliquity 35$^{\circ}$, without including the H$_2$O microphysical scheme (see Extended Data Fig. 2), shows a large depletion of water at mesospheric altitudes, resulting in a H escape up to two orders of magnitude lower than for the standard high obliquity simulation. This confirms that, as for current conditions (see Extended Data Fig. 1), the effects of radiative ice clouds and of supersaturation play a key role in the transport of water to the mesosphere and in the H escape, as suggested also by the close similarity between the seasonal variation of the middle atmosphere temperatures, the middle atmosphere water abundance, the hydrogen abundance at the exobase and the H escape. 
\\A sketch summarising the processes leading to larger H-escape in recent Mars is shown in Fig. \ref{fig:sketch}.

\subsubsection*{Discussion and limitations}

It is important to remind that this study is focused on the effects of increased obliquity only on the water cycle and escape, neglecting other atmospheric effects. An increased obliquity would likely imply many other changes in the planet atmosphere not captured in the current simulations that may in turn affect the escape. Incorporating some of these effects with very long characteristic temporal scales, such as the increased atmospheric pressure due to the larger polar insolation or the changes in atmospheric composition, would require integration times not currently achievable by GCMs. In addition, previous works using a similar strategy and simplifications as in this study \cite{Madeleine2014,Levrard2004,Forget2006} have proven successful to explain the mobilization of water at higher obliquity.

Another expected atmospheric effect at high obliquity is an increase in the dust lifting due to a reinforced meridional circulation \cite{Forget2017, Madeleine2009,Kurokawa2014},though the final effect on the atmospheric dust content is highly uncertain \cite{Forget2017}. In its current configuration the Mars-PCM requires prescription of a pre-established dust scenario designed to match observed or assumed dust opacities \cite{GonzalezGalindo2015}  (see Extended Data Fig. 3). Therefore, the model is unable to reliably simulate the effects of increased obliquity on the dust distribution in the atmosphere, and this is an important caveat.

In the simulations described above, a climatological dust scenario (labelled CLIM), appropriate for current Mars conditions, built on multi-annual dust climatology \cite{Montabone2015,Montabone2020} and representing a decadal average excluding Global Dust Storm years (GDS), was assumed for all obliquities. \iffalse The simulations shown in this study correspond to 3 consecutive Mars Years, started from initial files with stable water vapour abundances after running the model for about 50 Mars years (see also \ref{methods}). The instability of high obliquity simulation is another important caveat in this study, considering that longer simulations are required to obtain fully stable H-escape fluxes. Nevertheless, we estimated that annually integrated H-loss values from the last 2 consecutive MY differ by only 11\%  and 9\% for obliquity 30$^{\circ}$ and 35$^{\circ}$, respectively.\fi
It has also to be taken into account that H escape is intimately linked to O escape via a chemical feedback mechanism \cite{McElroy1972,Liu1976}. 
As also described in \cite{Chaffin2017,Jakosky2021}, in the long term this feedback mechanism would tend to decrease the increased H escape rate produced by increased obliquity until reaching equilibrium with O photochemical escape. In our simulations, we assume that the atmospheric composition in past conditions is similar to the current one, so this feedback mechanism is not considered, and so our calculated accumulated escape rate may be overestimated. Nevertheless, this chemical feedback process operates in temporal scales of million years \cite{Chaffin2017}, while changes in the obliquity have temporal scales of tens of thousand years \cite{Laskar2004}, suggesting that the ability of the feedback mechanism to reduce H escape on the long term is damped by the relatively faster changes in the obliquity.

\subsection*{Extrapolation of water loss back in time}

The escape flux can be expressed into a water Global Equivalent Layer (GEL), which is the thickness an amount of water would have if it were spread evenly over all Mars’ surface.  Assuming that present-day loss rates are representative of the H escape rate during the whole Mars history, we can extrapolate back in time to infer the time-integrated loss to space. The annually integrated H-loss for current obliquity is 1.03×10$^{34}$ atoms which translates into GEL values of 4.26 m during the last 4 billion years (see Table \ref{Tab:Rates_GEL}). This value is one order of magnitude smaller compared to the estimates for the total modern water inventory of about 3-25 m GEL (e.g. in the atmosphere and stored in the polar ice or subsurface ice) \cite{Scheller2021,Chaufray2015}. By contrast, simulations with obliquity of 35$^{\circ}$ give a GEL of 79 m, a factor of 18.52 higher than today. %This estimation increases to roughly 36 m if we consider a dustier atmosphere during high obliquity conditions. 

Our simulations are only relevant for the Amazonian period since early Mars was in a very different environment than the present one (e.g. with a fainter Sun at visible wavelength and stronger UV radiation). The limitations of our approach of assuming that the bulk atmosphere at different obliquities remains similar to the current one must also be taken into account, so the results should be interpreted as a first approximation to the full problem. Yet, those values are close to the lower limit of the range of geological estimates in early Mars history (100 to 1500 m GEL), and our study suggests that a substantial amount of water has escaped during high obliquity periods. H-escape has probably had a stronger role than that based on current estimates, leaving a lesser fraction to water sequestration in the crust.

\backmatter

\section*{Methods} \label{methods}

\subsection*{Recent model improvements}

The simulations shown in this paper were performed using an improved version of the ground-to-exosphere Mars Planet Climate Model \cite{Chaufray2021,Vals2022,Rossi2022, Gonzalez-Galindo2009}. Starting from a model version similar to that in \cite{Chaufray2021}, we have introduced different improvements in the model. 

First, in our simulations we added the sophisticated microphysical model of  \cite{Navarro2014} that considers ice nucleation on dust particles, ice particle growth, and scavenging of dust particles due to the condensation of ice. This scheme allows supersaturation in the atmosphere when there is a limited amount of cloud condensation nuclei (notably after scavenging) and/or when cooling rates are such that particle growth cannot condense all the supersaturated water. In comparison, \cite{Chaufray2021} assumed that water would instantaneously condense in cloud particles in order to keep the water concentration at or below saturation.
 Second, an improved photochemical scheme has been recently incorporated into the Mars-PCM \cite{Lefevre2021}, removing the need for using two different photochemical models in the lower and the upper atmosphere (see, e.g., \cite{Gonzalez-Galindo2013}). Within this new photochemical scheme, we have incorporated 4 additional ions (HCO$_{2}^{+}$, H$_{2}$O$^{+}$, H${_3}$O$^{+}$, and OH${^+}$) and 36 additional reactions, including the reactions of water vapour with the ionospheric species, which have been suggested to play an important role in the conversion of water to H in the upper atmosphere \cite{Stone2020}. In addition to the ionospheric reactions described in Table 2 in \cite{Gonzalez-Galindo2013}, the reactions in \ref{tab:H_2Oionreactions} have been incorporated into the model. The model version used is similar to that used in a recent study of the deuterium and hydrogen escape at aphelion conditions \cite{Chaufray2024}, and more details about the model can be found there.

Modelling the water cycle on Mars is challenging because of destabilising feedback, such as strong coupling with atmospheric temperature through clouds, and transport by global circulation. The water vapour mixing ratio in the mesosphere also depends on the supersaturation of the upper atmosphere, which is not well known. Furthermore, the effect of the microphysical processes at high altitude depends on model parameters not well constrained by observations. A notable recent improvement in the Mars-PCM is that the nucleation is now a function of temperature, which produces a more realistic supersaturation of water vapor in the upper atmosphere \cite{Naar2023}. All the processes involved in the water cycle and escape, such as the water sublimation/condensation cycle in the polar caps, transport, and penetration of water in the middle/upper atmospheres, chemical reactions transforming water into H were already included in the Mars-PCM, and discussed in detail in \cite{Chaufray2021,Lefevre2021,Naar2023}.

\subsection*{Factors affecting H escape}

In order to separate the effects of the different processes that have been included in the model, we show in  Extended Data Figure 1 the temporal evolution of the globally integrated H escape rate for three different simulations for current obliquity. The first simulation, shown in black, is a reference simulation using a similar physics as in \cite{Chaufray2021}, that is, the additional photochemical reactions for water-derived ions in  Supplementary Table 1 and the microphysical model have been switched off. In the second simulation, shown in blue, the microphysical model has been activated but the water-ion photochemical reactions remain off. And the third one, in red, is obtained with the full new model, that is, a simulation in which both the microphysical model and the additional H$_2$O ions photochemical reactions are included. To remove any possible influence of variations due to changes in the dust load or in the solar activity, we use in all three simulations an idealised scenario with a constant UV solar flux (only changing with the Sun-Mars distance), appropriate for solar average conditions, and a climatological dust scenario, appropriate for years without global dust storms. For a more quantitative comparison, the bottom panel of  Extended Data Figure 1  shows the ratio of the H escape rate obtained with the two last simulations to the reference simulation. 

The implementation of the microphysical model alone produces an increase of almost one order of magnitude in the H escape rate during the second half of the year with respect to the reference simulation. The reason is a much more permeable hygropause producing a strong increase in the water abundance in the mesosphere. As discussed in \cite{Vals2022} (see their Fig. 11), two main factors contribute to this strong modification of the water abundance: the radiative effects of water ice clouds, resulting in a warmer mesosphere, and the creation of supersaturated water layers significantly decreasing the water ice condensation. This result confirms the importance of the new scenario for hydrogen escape driven by the penetration of water to the mesosphere during a significant fraction of the Martian year. The impact of warm mesospheric temperatures in the penetration of water to the mesosphere during the 2018 GDS was already demonstrated using a different GCM \cite{Neary2020}. \cite{Shaposhnikov2019}, using another GCM, also showed that increased temperature during a GDS, combined with a reinforced meridional circulation, result in increased water vapour abundances above ~70 km.

On the other hand, during the aphelion season (Ls=60-150) the traditional scenario based on H$_2$ vertical diffusion seems to dominate, as the inclusion of the water microphysics does not increase the escape rate. In fact, the inclusion of the microphysical model produces a small decrease in the H escape rate during the aphelion season when compared to the reference simulation. The reason is that the microphysics significantly changes the simulated characteristics of the aphelion cloud belt \cite{Navarro2014}, which affects the atmospheric thermal structure, decreasing the temperature at the cloud level. This leads to a decrease of the water vapour pressure reducing the water abundances above the cloud level. 

The incorporation of the additional H$_2$O ions photochemical reactions further increases the simulated H escape rate in a factor generally between 10\% and 50\%, approximately, depending on the season. The reason is a more efficient conversion of the water accessing the thermosphere to H by means of the ionospheric reactions involving H$_2$O-derived ions, as previously found by \cite{Krasnopolsky2019,Stone2020}. Note however the larger effect during the Ls=270-300 period, when the escape rate increases by a factor of 2 compared to the simulation without extended chemistry. During this period water vapour penetrates to particularly high altitudes, reaching abundances of up to $\sim$30 ppm at the altitude of the ionospheric peak ($\sim$ 130 km, \cite{Withers2009}). This suggests that the scenario proposed by \cite{Montmessin2022} driven by the penetration of water to the mesosphere followed by its photolysis around 80 km and the subsequent vertical transport of hydrogen dominates the regular seasonal variation of hydrogen escape, while the water-ion chemistry scenario proposed by \cite{Chaffin2017} may play an important role in particular conditions where a more vigorous vertical transport could directly inject water vapour to altitudes where ionospheric reactions become effective. Our result aligns very well with recent findings combining different observational datasets and a 1-D photochemical model showing that water-ion chemistry becomes as important as water photolysis only during GDSs \cite{kleinbohl2024}.

The annually integrated H escape rate increases from 3.64×10$^{33}$ atoms for the reference simulation to 1.20×10$^{34}$ atoms for the simulation including water microphysics but not water ion chemistry and to 1.49×10$^{34}$ atoms for the simulation with both water microphysics and water ion chemistry. That is, the inclusion of microphysics produces an increase of a factor of about 4 in the annually integrated H escape rate, while the addition of H$_2$O ion chemistry further increases the integrated escape rate by $\approx$ 25\%. 

\subsection*{Role of global dust storms in current Mars}

To provide a quantitative estimation of the role of sporadic global dust storms with respect to the regular seasonal evolution for current obliquity, we performed simulations for three years affected by the presence of global dust storms (MY25, MY28, and MY34), replacing the observed day-to-day variability of the dust abundance as in \cite{Montabone2015,Montabone2020} by the climatological dust scenario described above, but keeping the observed day-to-day variability of the UV solar flux appropriate for each year. The results can be seen in Extended Data Figure 3. 
The effects of the global dust storm on H escape rates can be clearly seen after Ls $\approx$ 270$^{\circ}$ (MY28) and Ls $\approx$ 190$^{\circ}$ (MY25 and MY34). The increase in the instantaneous H escape rate at a given time during the storm can be as large as a factor of 4-5 approximately. Though limited in time, this increase also has a non-negligible effect over the annually integrated H escape rate, increasing it by roughly 50\% (MY25, from 1.62x10$^{34}$ to 2.41x10$^{34}$), 16\% (MY28, from 1.01x10$^{34}$ to 1.17x10$^{34}$) and 33\% (MY34, from 1.01x10$^{34}$ to 1.34x10$^{34}$). Our values are not far from the $\approx$18\% effect of the 2018 GDS over the accumulated escape rate recently found \cite{kleinbohl2024}. The differences between the three global dust storm years are related to the changes in the intensity of the storm (the storms in MY25 and MY34 being more intense than the one in MY28), and maybe also to the seasonality of the storm. It has to be noted, however, that the analysis of IUVS data obtained during the MY32-MY36 period did not show a significant increase in the seasonally averaged (Ls=180$^{\circ}$-360$^{\circ}$) escape rate during MY34 when compared to the other analysed years \cite{Mayyasi2023}. This may be due to the large averaging period and the complexity of deriving H escape rates from the IUVS Lyman-alpha spectra, involving a radiative transfer model to derive H exobase densities and an empirical fit to the exobase temperature measurements from NGIMS \cite{Mayyasi2023}, but it can not be ruled out that the model is overestimating the effects of the global dust storms on escape. Also, note the increase in H escape around Ls=330$^{\circ}$ in MY34, corresponding to the presence of a regional dust storm. The model predicts an increase in the instantaneous escape rate of about a factor of 2 to 3 with respect to the pre-storm values. This is in line with the analysis of the impact on the escape rate of this regional storm performed using data assimilation \cite{Holmes2021} and suggests that the effects of regional storms is well captured by the Mars-PCM.

\subsection*{Simulations for different obliquities}

In order to isolate the effects on the obliquity on the water cycle and escape, we intentionally kept the study simple, starting from a current Martian atmosphere and current water ice reservoirs, and modifying only the obliquity. Ground-to-exosphere simulations are computationally expensive, this is one of the main limitations of this work. Starting with initial files appropriate for current conditions, long simulations for high obliquity scenarios, and spanning several tens of Mars years, were performed with a version of the model limited up to about 150 km altitude. \iffalse, but not including all thermospheric processes. \fi The objective was to achieve a stable water cycle in the low-middle atmosphere, reducing the computational burden and the instabilities that arise with the full model extended up to the thermosphere. In these simulations, we also introduced recent modifications in the sub-stepping in the microphysical scheme that contributed to stabilize the simulations. Once the water cycle was stable, we run simulations for 3 consecutive Mars Years with the full model extended up to the thermosphere and including all physical and chemical processes relevant at those altitudes. The results shown in the paper are those obtained during the third year. 
For the obliquity 35$^{\circ}$ case the simulations show that a reasonable stability has been achieved (e.g. the difference between the beginning (Ls = 0) and the end Ls = 360$^{\circ}$ of the third Mars Years is $\sim$3\%)
%The annually averaged H escape rate for the obliquity 35$^{\circ}$ case changes in about ~10\% between year 2 and year 3} 

The horizontal resolution used in this work is 5.625$^{\circ}$ in longitude x 3.75$^{\circ}$, corresponding to grid boxes of the order of 330 x 220 km near the equator. The vertical resolution is variable, typically $\approx$ 10 km in the thermosphere. The scenario “climatology” (CLIM) is used here as reference for all obliquities, because it corresponds to our best guess for a typical Mars year. It was built using multiannual climatology and averaging MY24-MY31 dust scenarios (excluding data from both MY25 and MY28 GDS) as described in \cite{Montabone2015,Montabone2020}. More details can be found here:
\\https://www-mars.lmd.jussieu.fr/mars/dust$\_$climatology/index.html

\bmhead{Acknowledgements}

The IAA team (F.G.-G., G.G., M.A.L.-V., and A.B.) were funded by Spanish Ministerio de Ciencia, Innovación y Universidades, the Agencia Estatal de Investigación and EC FEDER funds under projects RTI2018-100920-J-I00, PGC2018-101836-B-100, and PID2022-137579NB-I00 and acknowledges financial support from the Severo Ochoa grant CEX2021-001131-S funded by MCIN/AEI/ 10.13039/501100011033. G.G. acknowledges financial support from Junta de Andalucía through the program EMERGIA 2021 (EMC21 00249). J.-Y.C. was partially funded by the Programme National de Planetologie of CNRS-INSU co-funded by CNES and Programme National Soleil Terre of CNRS-INSU co-funded by CNES and CEA. 
This project has received funding from the European Research Council (ERC) under the European Union’s Horizon 2020 research and innovation programme (grant agreement No 835275).

\bmhead{Author Contribution Statements}
G.G performed the simulations for high obliquity, analysed the results, discussed their implications, and wrote the first version of the paper. F.G.G conceived the study, contributed to the development and extension of the photochemical model and performed the simulations for current obliquity. J.-Y.C. contributed to the extension of the photochemical model and implemented the calculation of escape rates in MPCM. E.M. and F.F. are the main developers of MPCM and provided support for running the simulations. F.M., J.N., M.V. and L.R. contributed to improve the water cycle in the model. M.A.L.V.  and A.B contributed to validate the water cycle in the model. F.L. and Y.L. contributed to the development of the photochemical model and performed the spin-up for high-obliquity simulations. All the authors participated in the discussion of the results and contributed to the preparation of the manuscript.

\bmhead{Data availability}
The outputs of the MPCM used in this paper are available in NetCDF format at this link https://doi.org/10.5281/zenodo.15041471  \citep{Gilli2025}.

\bmhead{Code availability}
The Mars PCM is freely available from svn.lmd.jussieu.fr/Planeto/trunk.  For the high obliquity simulations the start files for obliquity 30$^\circ$ and obliquity 35$^\circ$ (in netCDF format) are also provided  at this link https://doi.org/10.5281/zenodo.15041471  \citep{Gilli2025}.

\bmhead{Competing Interests Statements}
The authors declare no competing interests.

%.%=========================
%
%.                   TABLES 
%
%===========================

%===============
%
%.  TABLE 1
% 
%==============

\begin{table}[!htbp]
%    \centering
    \resizebox{\columnwidth}{!}{%
    \begin{tabular}{|c|c|c|c|c|c|}
     \hline
     \hline
                     &  &  &  &  & \\     
                &  \textbf{H loss}  & \textbf{H loss}  & \textbf{H loss} &  \textbf{Ratio} &  \textbf{GEL (m)}   \\
                     &  &  &  &  & \\  
               \textbf{OBLIQUITY}&   (at/s) for 1 MY & (at/s) for 1 MY & (atoms)  & H$_{obl}$/H$_{current}$   &  during the last\\
                    &  &  &  &  & \\  
                & \textit{Average} & \textit{Min/Max}  & \textit{Annually} &  &  4 billion of years\\
                &  &  & \textit{integrated} &  & \\       
        
        \hline
        \hline
        &  &  &  &  & \\   
        \textbf{25.2$^{\circ}$} & \textbf{3.22 x 10$^{26}$} & \textbf{2.66 x 10$^{25}$/8.92 x 10$^{26}$} & \textbf{1.03 x 10$^{34}$} & \textbf{1} &  \textbf{4.26} \\
                   &  &  &  &  & \\       
         \hline
        &  &  &  &  & \\   
        \textcolor{orange}{\textbf{30$^{\circ}$}}  & \textcolor{orange}{\textbf{7.79 x 10$^{26}$}}  & \textcolor{orange}{\textbf{6.73 x 10$^{25}$/2.73 x 10$^{27}$}}  & \textcolor{orange}{\textbf{2.5 x 10$^{34}$}}  & \textcolor{orange}{\textbf{2.4}}  & \textcolor{orange}{\textbf{8.53}} \\
       &  &  &  &  & \\  
    \hline
     &  &  &  &  & \\  
     \textcolor{red}{\textbf{35$^{\circ}$}}  & \textcolor{red}{\textbf{6 x 10$^{27}$}}  & \textcolor{red}{\textbf{2.17 x 10$^{26}$/1.17 x 10$^{28}$}}  & \textcolor{red}{\textbf{1.9 x 10$^{35}$}}  & \textcolor{red}{\textbf{18.52}}  & \textcolor{red}{\textbf{79}} \\
      &  &  &  &  & \\  
     \hline
%      \hline
    \end{tabular}%
    }
    \caption{\textbf{Simulated H-loss for different obliquities and estimated Global Equivalent Layer (GEL) extrapolated back in time.} The annually integrated H escape obtained in our simulations range from 1.03x1034 (atoms) for our reference simulations in current obliquity  conditions, to 1.9x10$^{35}$ for obliquity 35$^{\circ}$ during the “CLIM” scenario. Assuming those values as representative of the H escape during the whole Mars history we can extrapolate the water loss back in time. The range of average GEL values that we obtained from annually integrated H-loss varies from 4.26 m for current obliquity to 79 m for obliquity 35$^{\circ}$. For comparison, the total water content measured in the reservoirs at the surface and subsurface is about 30 m, while the estimated amount of water needed to form the valley networks or outflow channels observed on Mars is of the order of 100s m GEL (with large uncertainties).}
    \label{Tab:Rates_GEL}
\end{table}

%%%%%%%%%%%%%%%%%%%%%%%%%%%%%%%%%%%%%%%%%%
%
%.    FIGURES
%
%%%%%%%%%%%%%%%%%%%%%%%%%%%%%%%%%%%%%%%%%%%

%%======= FIG. 1 =====================

%%======= FIG. 2  =====================

%%======= FIG. 3  =====================

%%======= FIG. 4  =====================

%%%%%%%%%%%%%%%%%%%%% 
%
%    BIBLIOGRAPHY
%
%%%

%\bibliography{bibliography}% common bib file

%\bibliography{sn-bibliography}% common bib file
%% if required, the content of .bbl file can be included here once bbl is generated
%%\input sn-article.bbl

%%%%%%%% ============>.  END DOCUNENT
\end{document}